\documentclass[showpacs,onecolumn,floats,superscriptaddress]{revtex4}
\usepackage{graphicx}

\begin{document}

\title{Surface effects in magnetic superconductors with a spiral magnetic
structure.}
\author{A.F. Volkov}

\affiliation {Institute f\"{u}r Theoretische Physik III,\\
Ruhr-Universit\"{a}t Bochum, D-44780 Bochum, Germany\\
and Institute for Radioengineering and Electronics of Russian Academy of\\
Sciences,11-7 Mokhovaya str., Moscow 125009, Russia\\}

\begin{abstract}
We consider a magnetic superconductor (\textit{MS}) with a spiral
magnetic structure. On the basis of generalized Eilenberger and
Usadel equations we show that near the boundary of the \textit{MS}
with an insulator or vacuum the condensate (Gor'kov's) Green's
functions are disturbed by boundary conditions and differ
essentially from their values in the bulk. Corrections to the bulk
quasiclassical Green's functions oscillate with the period of the
magnetic spiral, $2\pi /Q$, and decay inside the superconductor
over a
length of the order $v/2\pi T$ (ballistic limit) or $%
\sqrt{D/\pi T}$ (diffusive limit). We calculate the dc Josephson
current in an \textit{MS/I/MS} tunnel junction and show that the
critical Josephson current differs substantially from that
obtained with the help of the tunnel Hamiltonian method and bulk
Green's functions.
\end{abstract}

 \maketitle

\bigskip

\section{Introduction}

It is known that in some compounds the superconducting order can
coexist with a magnetic order of the ferromagnetic or
antiferromagnetic type. For example, in ternary rare-earth
compounds such as (RE)Rh$_{4}$B$_{4}$ and (RE)Mo$_{6}$X$_{8}$
(X=S,Se) the superconducting and magnetic ordering coexists in a
narrow temperature range (see the review \cite{BuzdinAdv} and a
more recent paper \cite{Buzdin01} and references therein). In
ErRh$_{4}$B$_{4}$ superconductivity takes place in the interval
0.7 $\leq $ $T$ $\leq $ 0.8 K, and the magnetic ordering arises
below $T_{m}$=0.8 K. In HoMo$_{6}$S$_{8}$ the magntic ordering
occurs below $T_{m}$= 0.74 K, whereas superconductivty exists in
the temperature range 0.7 $\leq $ $T$ $\leq $ 1.8 K. Besides, the
superconducting and magnetic order is realized in the layered
perovskite
ruthenocuprate compound RuSr$_{2}$GdCu$_{2}$O$_{8}$ \cite%
{Rut1,Rut2,Rut3,Rut4,Buzdin01}. In this compound an
antiferromagnetic order and, perhaps, a weak ferromagnetism take
place.

A uniform magnetization is impossible in a bulk superconductor as
the magnetic field destroys superconductivity. In order to explain
the coexistance of ferromagnetism and superconductivity, Ginsburg
and later Anderson and Suhl supposed that this coexistance is
possible in case of a domain or spiral magnetic structure
\cite{Ginsburg,Anderson}. The period of the magnetic structure has
been calculated in Ref.\cite{Anderson} (see also
Ref.\cite{Rusinov}), and on the order of magnitude it is equal to
$l_{m}\approx
2\pi (\xi _{0}k_{F})^{1/3}/k_{F}$, where $k_{F}$ is the Fermi momentum and $%
\xi _{0}=v_{F}/\pi \Delta _{0}$ is the correlation length in a clean
superconductor. For example, in HoMo$_{6}$S$_{8}$ the wave vector of the
periodic magnetic structure $Q\approx 0.03$ $A^{-1}$\cite{Bul83,Maple,Sinha}.

As is well known, many characteristics of a superconductor (the
critical temperature, the density-of-states etc) can be calculated
if the Green's functions of the system, including the anomalous
ones (or Gor'kov's functions), $\hat{F}$, are found \cite{AGD}.
These functions for a magnetic superconductor (\textit{MS}) with a
spiral structure have been obtained in Ref.\cite{Rusinov}. In this
case the functions $\hat{F}$ depend on the center-of-mass
coordinate and momentum direction so that the system is
anisotropic. Long ago, it was established that surface effects are
essential
for finite anisotropic samples such as anisotropic superconductors and high T%
$_{c}$ superconductors with $d$ wave pairing (see, for example, Ref.\cite%
{Buchholtz} and also the review \cite{Kirtley} and references
therein). In particular, the order parameter may be suppressed
near the
superconductor/vacuum or superconductor/insulator (\textit{S/V }or \textit{%
S/I)} interface. A high impurity concentration leads to averaging
the Green's functions in the momentum space so that in the
diffusive limit, characteristics of the system do not depend on
the sample size.

In this paper, we show that the surface effects are important in
\textit{MS}s with a spiral magnetic structure. In particular, the
Green's functions of the system are disturbed by boundary
conditions at the $S/V$ or $S/I$ interface in samples with any
impurity concentration. Corrections to the bulk Greens
functions due to boundary conditions oscillate in space with the period $%
2\pi /Q$ and decay from the interface over a length of the order $\xi
_{T}\approx v/2\pi T$ in the ballistic limit and of the order $\xi _{T}=%
\sqrt{D/\pi T}$ in the diffusive limit.

The surface effects become very important in the cases when one
needs to know the Green's functions near the interfaces. For
instance, the Josephson current $I_{J}$ in an \textit{MS/I/MS}
tunnel junction is
determined by the values of the Green's functions near the \textit{MS/I} interface (%
\textit{I }stands for an insulating layer). The Josephson current $I_{J}$ in
the \textit{MS/I/MS }junction with a spiral magnetic structure was
calculated in Ref.\cite{Kulic} on the basis of the tunnel Hamiltonian
method. The authors used the Gor'kov's functions calculated in Ref.\cite%
{Rusinov} for an infinite \textit{MS} with a spiral magnetic structure in
the ballistic limit. They have obtained that the Josephson critical current $%
I_{c}$ depends on the angle $\theta $ between the magnetization directions
in both \textit{MS}s near the interface and calculated the dependence of $%
I_{c}$ on different parameters of the junction (the exchange
field, the wave vector of the spiral, $Q$, etc). It has been
established that at some values of parameters
the critical current becomes negative ($\pi$ - state). We will show here that, although the current $%
I_{c}$ indeed depends on $\theta $ in a way similar to that in Ref.\cite%
{Kulic}, the dependence of $I_{c}$ on various parameters is
completely different. The point is that the tunnel Hamiltonian
method is not applicable to inhomogeneous superconductors and, in
particular, to \textit{MS}s with a spiral magnetization. In order to calculate $%
I_{c}$, one has to solve the Eilenberger or Usadel equation with
boundary conditions at the \textit{MS/I} interface. It turns out
that the Green's functions at the \textit{MS/I} interface differ
essentially from their values in the bulk, and correspondingly the
Josephson current also differs substantially from its value
obtained on the basis of the bulk Green's functions.

The structure of the paper is the following. In Sec. II, we
analyze the ballistic case. Using the Eilenberger equation
generalized for the case of the \textit{MS} with a magnetic
spiral, we find the spatial dependence of corrections to the bulk
Green's functions. In Sec. III, the diffusive case will be
considered. Using a generalized Usadel equation complemented by
boundary conditions at the \textit{MS/I} interface, we calculate
the Josephson current in \textit{MS/I/MS} tunnel junction and
compare the obtained critical Josephson current $I_{c}$ with that
obtained on the basis of the tunnel Hamiltonian method. In Sec.
IV, we discuss the obtained results.

\bigskip

\section{Ballistic case}

We consider a \textit{MS} with a spiral magnetic structure. The
exchange field acting on free electrons is assumed to lay in the
$(y,z)$ plane and to rotate in space with the wave vector $Q$;
that is, the vector of the exchange field is: $\mathbf{h}=h(0,\sin
\alpha (x),\cos \alpha (x))$ with
$\alpha =Qx+\theta, x\geq 0$, ($\theta $ is the angle between the magnetization and $z$%
-axis at $x=0$). The superconducting order parameter $\Delta $ is
taken into account in the mean field approximation: $\Delta
=\lambda _{S}\sum_{p}\langle \psi _{\uparrow ,p}\psi _{\downarrow
,-p}\rangle $, i.e. the singlet pairing is assumed. The
Eilenberger equation is derived in a standard way (see, for
example, \cite{Eilen,BVErmp,Buzdinrmp,LO,Kopnin}). The main
difference between the cases of an ordinary, nonmagnetic
superconductor and \textit{MS} with a spiral structure is that the
quasiclassical Green's function $\check{g}$ in the latter case is
a $4\times 4$ matrix in the Gor'kov-Nambu and spin space. This
equation has the form

\begin{equation}
i\textbf{v}\nabla \check{g}+\omega \lbrack \hat{\tau}_{3}\otimes \hat{\sigma}_{0},%
\check{g}]+i[\mathbf{h(}x\mathbf{)S},\check{g}]+\left[ \hat{\Delta}\otimes
\hat{\sigma}_{3},\check{g}\right] +(i/2\tau )\left[ \langle \check{g}\rangle
,\check{g}\right] =0\;.  \label{Eilen}
\end{equation}
where $v$ is the Fermi velocity, $\mathbf{S=(\hat{\sigma}_{1},\hat{\sigma}%
_{2},\hat{\tau}_{3}\otimes \hat{\sigma}_{3}),}$
$\hat{\sigma}_{k},\hat{\tau}_{k}$ are the Pauli matrices in the
spin and Gor'kov-Nambu space, and $\hat{\sigma}_{0},
\hat{\tau}_{0}$ are the unit matrices. The square and angle
brackets mean the commutator and averaging over angles,
respectively, and $\tau $ is an elastic scattering time. In order
to exclude the coordinate dependence of the third term in
Eq.(\ref{Eilen}), we perform a transformation (see Ref.
\cite{BVErmp})

\begin{equation}
\check{g}=\check{U}\mathbf{\otimes }\check{g}_{n}\mathbf{\otimes }\check{U}%
^{+},  \label{rotation}
\end{equation}
where $\check{U}=\hat{\tau}_{0}\otimes \hat{\sigma}_{0}\cos (\alpha
/2)+i\sin (\alpha /2)\hat{\tau}_{3}\otimes \hat{\sigma}_{1}$ is an operator
corresponding to a rotation in the spin and particle-hole space, and $\check{%
g}_{n}$ is a new matrix. Then Eq.(\ref{Eilen}) acquires the form

\begin{equation}
v\mu \partial _{x}\check{g}+[\hat{\tau}_{3}\otimes (\omega \hat{\sigma}%
_{0}+ih\hat{\sigma}_{3},\check{g}]+iv\mu(Q/2)[\hat{\tau}_{3}\otimes \hat{\sigma}%
_{1},\check{g}]+\left[ \hat{\Delta}\otimes
\hat{\sigma}_{3},\check{g}\right] +(i/2\tau )\left[ \langle
\check{g}\rangle ,\check{g}\right] =0\; \label{EilenModif}
\end{equation}
where $\mu =p_{x}/p$. The subindex ''$n$'' is omitted. \ From the physical
point of view, the transformation given by Eq.(\ref{rotation}) means the
transition to a rotating coordinate system, in which the magnetization
vector is directed along the $z$-axis. That is why the exchange field $h$ in
Eq.(\ref{EilenModif}) contains only the $z$-component.

For simplicity, we restrict the consideration with the case of
temperatures close to the critical one of the superconducting
transition, $T_{c}$. In this case the matrix Green's function
$\check{g}$ may be represented in the form

\begin{equation}
\check{g}=sign\omega \cdot \hat{\tau}_{3}\otimes \hat{\sigma}_{0}+\check{f}%
\;.  \label{Eq2}
\end{equation}
where the anomalous (Gor'kov's) matrix function, $\ \check{f}\;$,
is assumed to be small, that is, all elements of this matrix are
small. The first term is the normal, matrix Green's function in
the Matsubara representation.

In this Section, we consider the ballistic case, i.e. we suppose that $\tau
\rightarrow \infty $. Substituting the matrix $\check{g}$ from Eq.(\ref{Eq2}%
) into Eq.(\ref{Eilen}), we come to the equation for the anomalous function $%
\check{f}\;$

\begin{equation}
v\mu \hat{\tau}_{3}\mathbf{\otimes }\partial _{x}\check{f}\;+iv\mu (Q/2)[%
\hat{\sigma}_{1},\check{f}]_{+}+2\omega \check{f}\;+ih[\hat{\sigma}_{3},%
\check{f}]_{+}=\hat{\tau}_{2}\otimes \hat{\sigma}_{3}\Delta sign\omega .
\label{Eilen3}
\end{equation}

We represent the matrix $\check{f}\;$in the form

\begin{equation}
\check{f}\;=\hat{f}\otimes \hat{\tau}_{2}+\hat{F}\otimes \hat{\tau}_{1}
\label{fRepresent}
\end{equation}
where $\hat{f}$ and $\hat{F}$ are matrices in the spin space that can be
represented as a sum of Pauli matrices

\begin{equation}
\hat{f}=\sum_{k}f_{k}\hat{\sigma}_{k};\ \hat{F}=\sum_{k}F_{k}\hat{\sigma}_{k}
\label{fFPauli}
\end{equation}
where $k=0,1,3.$

Eq.(\ref{Eilen3}) is a system of linear equations with respect to
coefficients $f_{k}$ and $F_{k}$. The solution of these equations
consists of a part, $\bar{f}_{k}$ and $\bar{F}_{k}$, constant in
space and a nonhomogeneous part, $\delta f_{k}(x)$ and $\delta
F_{k}(x).$ The latter part arises if there are nontrivial boundary
conditions in the problem. The homogeneous part is a solution for
an infinite sample when boundary conditions can be ignored. The
homogeneous solution can be easily found. It has the form

\begin{equation}
\bar{f}_{3}=\frac{\Delta (\epsilon _{Q}^{2}+\omega ^{2})}{|\omega
|(\epsilon _{Q}^{2}+h^{2}+\omega ^{2})},\text{ \
}\bar{f}_{0}=-\frac{ih\Delta sign\omega }{|\omega |(\epsilon
_{Q}^{2}+h^{2}+\omega ^{2})} \label{HomSolution}
\end{equation}
where $\epsilon _{Q}=\mu vQ/2$. All other coefficients (i.e.
$f_{1},F_{k}$) equal to zero. The coefficient $\bar{f}_{3}$ is the
amplitude of the singlet component, and the coefficient
$\bar{f}_{0}$ is the amplitude of the triplet component with zero
projection of the total spin of a Cooper pair on the $z$-axis (in
the rotating coordinate system), $S_{z}=0$. The singlet component
is an even function of $\omega ,$ while the triplet component,
$\bar{f}_{0}$, is an odd function of $\omega $ \cite{BVErmp}. One
can see that the exchange field, $h$, suppresses the amplitude
$\bar{f}_{3}$, whereas at a sufficiently large wave vector of the
spiral $Q$, the amplitude $\bar{f}_{3}$ is restored to the value
$\Delta /|\omega |$ which is the amplitude of the condensate
function in a nonmagnetic superconductor. Note that the authors of
Ref.\cite{Kulic} used only bulk solutions in the laboratory
coordinate frame. These functions may be reduced to the
quasiclassical Green's functions in Eq.(\ref{HomSolution}).

The function $\bar{f}_{3}$ determines a change of the critical
temperature of the superconducting transition, $T_{c}$, due to the
exchange field $h$ and wave vector of the magnetic spiral $Q$
(see, for example, the review articles \cite{Buzdinrmp,BVErmp})

\begin{equation}
\frac{T_{c0}-T_{c}}{T_{c}}=2\pi T\sum_{\omega }\int_{0}^{1}d\mu \lbrack \frac{1}{%
|\omega |}-\frac{\bar{f}_{3}}{\Delta }]=2\pi T\sum_{\omega }\int_{0}^{1}d\mu \frac{%
h^{2}}{|\omega |(\epsilon _{Q}^{2}+h^{2}+\omega ^{2})},\text{ }
\label{Tc}
\end{equation}
where $T_{c0}$ is the critical temperature in the absence of the exchange
field $h$. It is seen that with decreasing the spiral period, $2\pi /Q,$ the
suppression of the critical temperature is reduced and at $vQ>>h$ the
critical temperature is the same as in a nonmagnetic superconductor, i.e. $%
T_{c}\rightarrow T_{c0}.$

Now we turn to the calculation of corrections $\delta f_{k}$ and
$\delta F_{k}$ that arise due to boundary conditions and depend on
$x$. Note that if the correction $\delta f_{3}$ is not small
compared to $\bar{f}_{3}$, a correction to the order parameter
$\delta \Delta (x)$ will not be small as
well. This circumstance makes the problem rather complicated because Eq.(\ref%
{Eilen3}) becomes a system of six equations with the right-hand
side which depends on $x$. In order to simplify the problem, we
assume that the correction $\delta f_{3}$ is small and we can
neglect a variation of $\Delta $ in space. We will see below that
in a general case $\delta f_{3}$ may be comparable with
$\bar{f}_{3}$. In this case our results are correct up to a
numerical factor of the order unity. In the next Section, we
discuss the validity of the obtained results in more detail.

Thus, in order to find the corrections $\delta f_{k}$ and $\delta
F_{k},$ \ we have to solve a system of homogeneous linear
equations (\ref{Eilen3}) without\ the right-hand side.
Substituting the expansions (\ref{fFPauli}) with $\delta f_{k}$
and $\delta
F_{k}$ as the coefficients of these expansions into Eq.(\ref{Eilen3}) \ with $%
\delta \Delta =0$ and \ representing the coordinate dependence of
these coefficients in the form $\{\delta f_{k}$,$\delta
F_{k}\}\sim \exp (\kappa x) $, we obtain a system of six linear
equations. One can see from these equations that the coefficients
$f_{1}$ and $F_{0,3}$ are antisymmetric functions of $\mu $,
whereas the coefficients $f_{0,3}$ and $F_{1}$ are symmetric
functions of $\mu $. We do not write down these equations as they
are rather cumbersome. Instead of this, we write the determinant
of the system which determines the eigenvalues $\kappa _{i}.$ It
is reduced to a cubic algebraic equation

\begin{equation}
\lambda _{\omega }(1-z)[\lambda _{\omega }(1-\lambda _{\omega
})+z^{2}]+(1-\lambda _{\omega })(1+z)^{2}(\lambda _{\omega }-z)=0
\label{Det}
\end{equation}
where $z=\epsilon _{\kappa }^{2}/\Omega ^{2},$ $\lambda _{h}=h^{2}/\Omega
^{2},$ $\lambda _{Q}=\epsilon _{Q}^{2}/\Omega ^{2},$ $\Omega ^{2}=\omega
^{2}+h^{2}+\epsilon _{Q}^{2},$ $\epsilon _{\kappa }=\mu v\kappa /2,$and $%
\epsilon _{Q}=\mu vQ/2$.

In order to find the eigenvalues, one has to solve this equation.
We
consider the most interesting case of large energy $\epsilon _{Q}:$ $%
\epsilon _{Q}>>T,h$. In this case the critical temperature $T_{c}$ is close
to $T_{c0}$. The solutions of Eq.(\ref{Det}) are

\begin{equation}
z_{1}\cong \lambda _{\omega },\text{ }\kappa _{1}\cong 2|\omega |/(v|\mu |)
\label{z1}
\end{equation}
and

\begin{equation}
z_{2,3}\cong -1\pm 2i\sqrt{\lambda _{\omega }},\text{ }\kappa _{2,3}\cong
\pm iQ-2|\omega |/(v|\mu |)  \label{z23}
\end{equation}

Therefore, the eigenfunctions corresponding to $\kappa _{2,3}$
oscillate in space with the period of the spiral and decay over
the distance of the order $\xi _{T}=v/2\pi T.$ The eigenfunction,
which corresponds to $\kappa _{1}$, decreases monotonously from
the interface over the correlation length $\xi _{T}.$

The amplitudes $f_{k}$ and $F_{k}$ may be found from boundary conditions at
the \textit{MS/V} or \textit{MS/I} interfaces \cite{Zaitsev}

\begin{equation}
\check{f}\;(\mu )-\check{f}\;(-\mu )=0  \label{BC}
\end{equation}
which read that the antisymmetric part of the Green's function
should turn to zero at the \textit{MS/I} interface. One can solve
the corresponding equations and find the amplitudes $f_{k}$ and
$F_{k}$. However we will not do that for two reasons. First, the
corresponding expressions are cumbersome. The second and more
important reason is that the surface effects are displayed near
the interface at which a random (diffusive) scattering takes
place. Therefore, the ballistic case considered in this Section is
not relevant to this situation. In the next Section we consider a
more realistic case of a sample with a high impurity concentration
(dirty case). We will find the eigenvalues $\kappa _{i}$ and the
amplitudes of eigenfunctions. One can show that the structure and
form of the dependencies of the functions $f_{k}$ and $F_{k}$ on
$\mu $ and $\omega $ are qualitatively the same in both cases,
ballistic and diffusive. The only difference is that, whereas in
the diffusive case only the zero and first terms in the expansion
in spherical harmonics are important, in the ballistic case the
dependence on $\mu $ is more complicated.

\bigskip

\section{Diffusive case}

\bigskip In this Section, we consider the influence of the boundary on the
condensate functions assuming that the impurity concentration is
high and the condition $l<<{2\pi /Q,\xi _{T}}$ is satisfied, where
$l=v\tau $ is the mean free path. In this case the part of the
condensate function $\check{f}\;_{asm}$ antisymmetric in the
momentum space is expressed through the symmetric part via the
well known expression \cite{Usadel,Buzdinrmp,BVErmp,LO,Kopnin}

\begin{equation}
\check{f}\;_{asm}=(p_{x}/|p|)\hat{\tau}_{3}sgn\omega (\partial _{x}\check{f}\;+i(Q/2)%
\hat{\tau}_{3}[\hat{\sigma}_{1},\check{f}]_{+})  \label{Asym}
\end{equation}%
where $\hat{\tau}_{3}\mathit{sgn}\omega $ is the ordinary
quasiclassical Green's function in the normal state (see
Eq.(\ref{Eq2})). The second term arises as a result of the
transformation (\ref{rotation}), the term
$[\hat{\sigma}_{1},\check{f}]_{+}$ means anti-commutator. One can
see that the asymmetric part has the opposite parity in $\omega $
compared to the symmetric part $\check{f}\;$; if $\check{f}\;$\ is
an odd function of $ \omega $, then $\check{f}\;_{asm}$ is an even
function of $\omega $ and vice versa \cite{BVErmp}. In the
simplest case of ordinary BCS superconductors the symmetric
function near $T_{c}$ is equal to $\check{f}=\hat{\tau}_{3}\otimes
\hat{\sigma}_{0}\Delta /|\omega |$, i.e. is an even function of
$\omega$. Obviously the antisymmetric part $\check{f}\;_{asm}$ is
an odd function of $\omega$. This issue is discussed in detail in
Refs.\cite{Golubov}.

We assume again that the temperature is close to $%
T_{c}$. The symmetric part of the condensate function $\check{f}\;$\ after
the transformation Eq.(\ref{rotation}) obeys the equation \cite{BVErmp}

\begin{equation}
D\{-\partial _{xx}^{2}\check{f}+\frac{Q^{2}}{2}(\check{f}+\hat{\sigma}%
_{1}\otimes \check{f}\otimes \hat{\sigma}_{1})\;+i\frac{Q}{2}\hat{\tau}_{3}[%
\hat{\sigma}_{1},\partial _{x}\check{f}]_{+}\}+2|\omega |\check{f}%
\;+ih_{\omega }[\hat{\sigma}_{3},\check{f}]_{+}=2\hat{\tau}_{2}\otimes \hat{%
\sigma}_{3}\Delta .  \label{Usadel}
\end{equation}
Here $h_{\omega }=\mathit{sgn}\omega h.$ As follows from
Eqs.(\ref{BC},\ref {Asym}) the boundary condition has the form

\begin{equation}
\partial _{x}\check{f}\;+i(Q/2)\hat{\tau}_{3}[\hat{\sigma}_{1},\check{f}%
]_{+}=0  \label{BCdif}
\end{equation}
This means that the spiral axis is assumed to be perpendicular to the
\textit{MS/V} or \textit{MS/I} interface.

One can see that a coordinate-independent solution for
Eq.(\ref{Usadel}) satisfies the boundary condition only if $Q=0$.
If $Q$ is not zero, the anti-commutator
$[\hat{\sigma}_{3},\bar{f}_{0}\hat{\sigma}_{0}]_{+}\neq 0$, and
therefore $\partial _{x}\check{f}$ also differs from zero at the
boundary.

We have to solve Eq.(\ref{Usadel}) with the boundary condition (\ref{BCdif}%
). The uniform solution again has the form (\ref{HomSolution}) with $%
\epsilon _{Q}=DQ^{2}/2.$ The correction $\delta \check{f}\;=\check{f}\;-\hat{%
\tau}_{2}(\bar{f}_{3}\hat{\sigma}_{3}+\bar{f}_{0}\hat{\sigma}_{0})$
satisfies the uniform equation (\ref{Usadel}) and may be
represented in the form (\ref{fRepresent}-\ref{fFPauli}), where
only the coefficients $f_{0,3}$ and $F_{1\text{ }}$differ from
zero, that is

\begin{equation}
\check{f}\;=(f_{3}\hat{\sigma}_{3}+f_{0}\hat{\sigma}_{0})\otimes \hat{\tau}%
_{2}+F_{1}\hat{\sigma}_{1}\otimes \hat{\tau}_{1}  \label{fDif}
\end{equation}

We look for a solution in the form of exponentially decaying functions: $%
\delta \check{f}\;\sim \exp (\kappa x)$ with $\text{Re}\kappa <0.$
The determinant of the system of Eqs.(\ref{Usadel}) has the form

\begin{equation}
\lbrack (1+z)^{2}+2\lambda _{\omega }(1-z)+\lambda _{\omega }^{2}](\lambda
_{\omega }-z)+\lambda _{h}^{2}(1+\lambda _{\omega }-z)=0  \label{Detdif}
\end{equation}
where $z=(\kappa /Q)^{2},\lambda _{\omega }=2|\omega |/DQ^{2},\lambda
_{h}=2h_{\omega }/DQ^{2}$.

Again we consider the most interesting case of small $\lambda
_{\omega ,h}$ which seems to be relevant to the experiment
\cite{Maple}: $\{\lambda _{\omega },\lambda _{h}\}<<1,$ i.e.
$\{T,h\}<<DQ^{2}$ \cite{footnote}. In this limit the eigenvalues
are

\begin{equation}
z_{1}=\lambda _{\omega }+\lambda _{h}^{2};\text{ }\kappa _{1}=-Q\sqrt{%
\lambda _{\omega }+\lambda _{h}^{2}}  \label{z1}
\end{equation}
and

\begin{equation}
z_{2,3}=-1\pm i\sqrt{2(2\lambda _{\omega }+\lambda _{h}^{2})};\text{ }\kappa
_{2,3}=\pm iQ-Q\sqrt{(2\lambda _{\omega }+\lambda _{h}^{2})/2}  \label{z23}
\end{equation}

Thus, the correction $\delta f_{3}$ may be written as

\begin{equation}
\delta f_{3}(x)=a_{1}\exp (\kappa _{1}x)+a_{+}\exp (\kappa _{+}x)+a_{-}\exp
(\kappa _{-}x)  \label{f3}
\end{equation}
where $\kappa _{+}=\kappa _{2}=+iQ-Q\sqrt{(2\lambda _{\omega }+\lambda
_{h}^{2})/2}$ and $\kappa _{-}=\kappa _{3}=\kappa _{+}^{\ast }$. The first
term decreases monotonously inside the superconductor, whereas the second
and third terms oscillate with the period $2\pi /Q$ and decay over the
length of the order of $\min \{\xi _{T},\sqrt{(DQ^{2}/h)(D/h)}\}$. The
corrections, $\delta f_{0}$ and $F_{1},$ have the form

\begin{equation}
\delta f_{0}(x)=-i\lambda _{h}a_{1}\exp (\kappa _{1}x)-(i\lambda
_{h})^{-1}[(1+i\alpha )a_{+}\exp (\kappa _{+}x)+(1-i\alpha )a_{-}\exp
(\kappa _{-}x)]  \label{f0}
\end{equation}

\begin{equation}
F_{1}(x)=2i\lambda _{h}\sqrt{z_{1}}a_{1}\exp (\kappa _{1}x)-\lambda
_{h}^{-1}[(1+i\alpha )a_{+}\exp (\kappa _{+}x)-(1-i\alpha )a_{-}\exp (\kappa
_{-}x)]  \label{F1}
\end{equation}
where $\alpha =\sqrt{2(2\lambda _{\omega }+\lambda _{h}^{2})}$. The
coefficients $a_{1}$ and $a_{\pm }$ are found from the boundary condition (%
\ref{BCdif})

\begin{equation}
a_{1}=-\frac{i\lambda _{h}}{\lambda _{h}^{2}+\alpha \sqrt{z_{1}}/2}\bar{f}%
_{0},\text{ }a_{+}=a_{-}^{\ast
}=-ia_{1}\sqrt{z_{1}}[\frac{1}{2}-i\frac{\lambda _{h}^{2}+\alpha
^{2}/2}{\alpha }]  \label{a1}
\end{equation}

Making use of Eqs.(\ref{f3}-\ref{F1}), one can obtain the values
of the condensate function at the interface $\check{f}(0)$ that
determine, for example, the Josephson current in \textit{MS/I/MS}
junction. We find

\begin{equation}
f_{3}(0)\cong -\frac{(\alpha /2)\sqrt{z_{1}}}{(\lambda
_{h}^{2}+\alpha \sqrt{z_{1}}/2)}\bar{f}_{3},\text{ } f_{0}(0)\cong
-i\lambda _{h}\frac{\lambda _{\omega }-\lambda _{h}^{2}/2}{\lambda
_{\omega }+\lambda _{h}^{2}/2}f_{3}(0), \label{Corrdif}
\end{equation}
and
\begin{equation}
\ F_{1}(0)=-i \frac{2\lambda _{h}}{\alpha }f_{3}(0)
\label{Corr1dif}
\end{equation}
where the amplitude of the bulk singlet $\bar{f}_{3}$ component can be
expressed in terms of the parameters $\lambda _{\omega ,h}$

\begin{equation}
\bar{f}_{3}=\frac{(\lambda _{\omega }+1)}{\lambda _{\omega
}(\lambda _{\omega }+1)+\lambda_{h}^{2}}\frac{2\Delta }{DQ^{2}},
\label{fBar}
\end{equation}

\begin{figure}
\begin{center}
\includegraphics[width=0.6\textwidth]{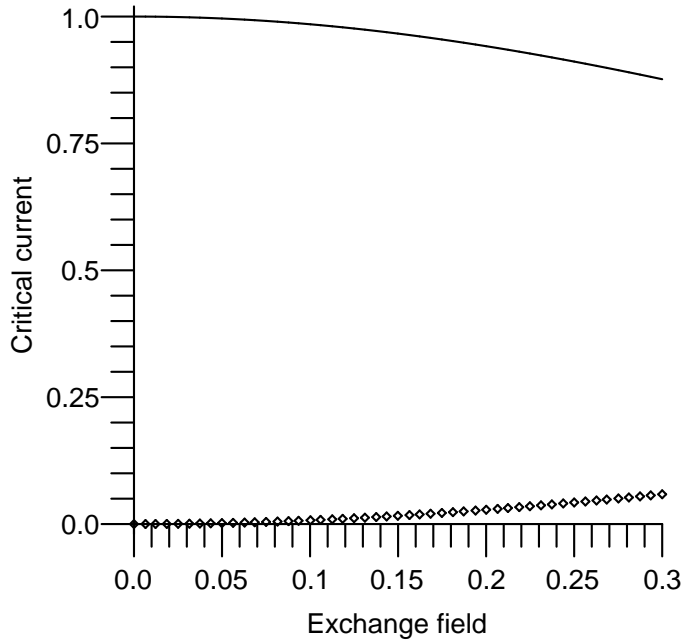}
\end{center}
\large{\caption{Contributions of the singlet (solid line) and
triplet $S_{z}=0$ (dotted line) components to the Josephson
critical current as a function of the exchange field. This
dependence has been obtained on the basis of bulk quasiclassical
Green's functions and corresponds to the tunnel Hamiltonian
method. The critical current $I_{c}(h)$ and the exchange field $h$
are plotted in units $I_{c}(0)$ and $DQ^{2}/2$, respectively. The
temperature is chosen equal to $T=0.1DQ^{2}/2\pi $.}}
\end{figure}

In the considered limit, $\lambda _{\omega ,h}\ll 1,$ the function $\bar{f}%
_{3}$ is close to the value of the singlet component in an
ordinary (nonmagnetic) superconductor. The exchange field, which
tries to destroy Cooper pairs, is effectively averaged due to
rotation of the magnetization vector.

Now we discuss the conditions under which the obtained results are
valid. Consider first the case of a thick sample ($d\gg \xi
_{GL}\cong 1.2\sqrt{D/T}(T/\Delta )$, where $d$ is the thickness
of the sample and $\xi _{GL}$ is the Ginsburg-Landau correlation
length)
One can see that if $\lambda _{h}^{2}>>\lambda _{\omega }(n=0)$, i.e. $%
h^{2}>>(\pi T)DQ^{2},$ the value of $f_{3}(0)$ is $f_{3}(0)=\bar{f}_{3}\sqrt{%
2}/(2+\sqrt{2})\approx 0.41\bar{f}_{3},$ i.e. the singlet
condensate function at the interface differs from the bulk value
by a numerical factor of the order $1$. In the limit $\lambda
_{h}^{2}<<\lambda _{\omega }$ the
singlet component is almost constant in space so that $f_{3}(0)\approx \bar{f%
}_{3}$. Therefore our results are valid in this limit. However,
our results are also correct if the thickness of the sample $d$ is
less than the Ginsburg-Landau correlation length. In this case,
the order parameter $\Delta$ is constant in space
\cite{Ovchinnikov} so that our assumption about the
coordinate-independent $\Delta$ is fulfilled and the obtained
results are exact.

\begin{figure}
\begin{center}
\includegraphics[width=0.6\textwidth]{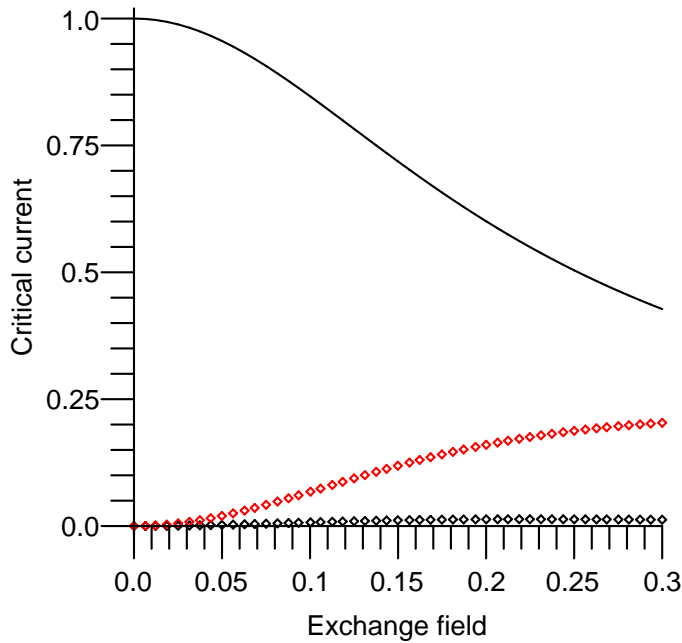}
\end{center}
\large{\caption{Contributions of the singlet (solid line) and
triplet (dotted lines) components to the Josephson critical
current as a function of the exchange field $h$. The upper and
lower dotted curves correspond to the triplet $|S_{z}|=1$ and
$S_{z}=0$ condensate components. These curves are plotted on the
basis of Eq.(\ref{JosCurrent}). The normalization units are the
same as in Fig.1.}}
\end{figure}

Let us discuss the meaning of the component of the condensate function $%
\check{f}\;$. As we said above, the function $f_{3}(0)$ is the
amplitude of the singlet component at the interface. The function
$f_{0}(0)$ is the amplitude of the triplet $S_{z}=0$ component.
One can see that both functions, $\bar{f}_{0}$ and $\delta f_{0}$,
(the bulk value and the correction due to the surface effects) are
small compared to the singlet value in the considered limits,
$\lambda _{\omega ,h}<<1.$ The function $F_{1}$ is the amplitude
of the triplet component with $|S_{z}|=1$ in the rotating
coordinate system. In the bulk, it is equal to zero. Just this
component penetrates the ferromagnet over a long distance in $S/F$
structures with a rotating magnetization
\cite{BVErmp,Sosnin,VAE,FVE,Eschrig}. This triplet component
$F_{1}(0)$ is of the order of the singlet component in the bulk,
$\bar{f}_{3},\ $ at $\lambda _{\omega }<<\lambda _{h}^{2}$ and
less than $\bar{f}_{3}$ at $\lambda _{\omega }>>\lambda _{h}^{2}$.

\begin{figure}
\begin{center}
\includegraphics[width=0.6\textwidth]{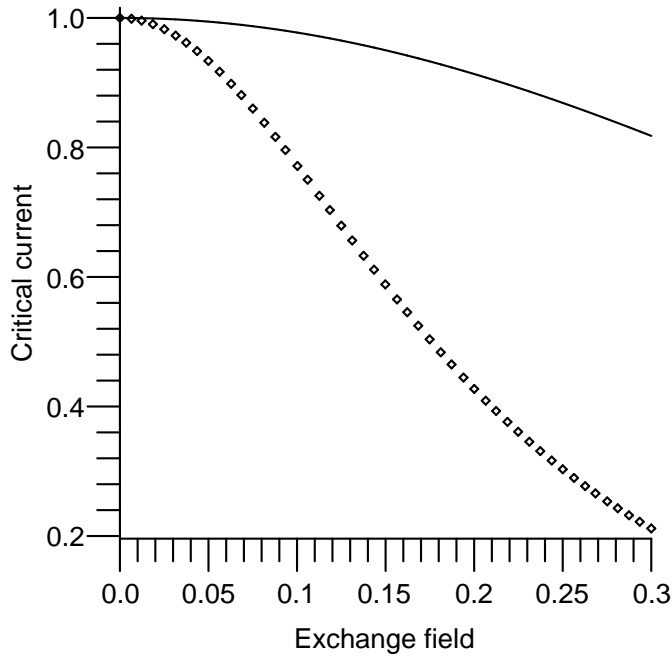}
\end{center}
\large{\caption{The total critical current versus the exchange
field obtained on the basis of the bulk Green's functions (solid
line) and of the Green's functions taken at the \textit{MS/I}
interface (dotted line). The solid line corresponds to the tunnel
Hamiltonian method. The normalization units are the same as in
Fig.1.}}
\end{figure}

Knowing the quasicalssical Green's functions at the \textit{MS/I}
interface, we can calculate the dc Josephson current $I_{J}$ in a
\textit{MS/I/MS} tunnel junction consisiting of two \textit{MSs}.
The Josephson current in this junction is expressed in terms of
the components $f_{0,3}$ and $F_{1}$ at the interfaces
\textit{MS/I}$,i.e.$ at $x=0$ (see the Appendix)

\begin{equation}
I_{J}=I_{c}\sin \varphi ,\text{ \ }I_{c}=(eR_{B})^{-1}(2\pi T)\sum_{\omega
=0}^{\infty }(f_{3}^{2}(0)+\cos \theta \lbrack f_{0}^{2}(0)+F_{1}^{2}(0)])
\label{JosCurrent}
\end{equation}
where $R_{B}$ is the resistance of the junction in the normal state, $%
\varphi $ is the phase difference, $\omega =\pi T(2n+1)$ is the
Matsubara frequency, and $\theta $ is the angle between the
magnetization vectors in the right and left magnetic
superconductors at the interfaces. Since we are
interested in the Josephson current in the lowest order in the parameter $%
R_{B}^{-1}$, the functions $f_{0,3}(x)$ and $F_{1}(x)$ should obey the
boundary conditions (\ref{BCdif}) that correspond to the limit $%
R_{B}\rightarrow \infty $. These functions are given by
Eqs.(\ref{Corrdif}-\ref{Corr1dif}).

A formula, which resembles Eq.(\ref{JosCurrent}), was obtained in
Ref.\cite{Kulic} on the basis of the tunnel Hamiltonian method.
What is the difference between these two formulae? First, the term
$F_{1}^{2}(0)$ is absent in Ref.\cite{Kulic}. Second, instead of
terms $f_{0,3}^{2}(0),$ in Ref.\cite{Kulic} there are terms
$\bar{f}_{0,3}^{2}$ corresponding to the bulk solutions. This
difference leads to essential consequences. In particular, the
conclusion made in Ref.\cite{Kulic} about the possibility to
realize a $\pi -$junction for some values of parameters such as
$h,Q$ etc is not justified.

Fig.1 shows the contributions of the bulk singlet ($\bar {f}_{3}$)
and $S_{z}=0$ triplet ($\bar {f}_{0}$) components to the critical
current and corresponds to the tunnel Hamiltonian method. Fig.2
displays the contributions of the singlet ($f_{3}$), $S_{z}=0$
triplet ($f_{0}$) and $|S_{z}|=1$ triplet ($F_{1}$) components to
the critical current. The solid and dotted curves in Fig.2 are
normalized partial critical currents defined as

\begin{equation}
\text{ \ }i_{3}=\frac{8T^{2}}{\pi ^{2}\Delta ^{2}}\sum_{\omega }f_{3}^{2}(0),%
\text{ }i_{0}=-\frac{8T^{2}}{\pi ^{2}\Delta ^{2}}\sum_{\omega }f_{0}^{2}(0),%
\text{ \ }i_{1}=-\frac{8T^{2}}{\pi ^{2}\Delta ^{2}}\sum_{\omega }F_{1}^{2}(0)
\label{JosCurrents103}
\end{equation}
where $\pi ^{2}/8=\sum_{\omega }\left( 2n+1\right) ^{-2}$ is the
normalization factor. The functions $f_{3,0}^{2}(0)$ and
$F_{1}(0)$ are given by Eqs.(\ref{Corrdif}-\ref{Corr1dif}). The
lower (upper) dotted lines are due to the $S_{z}=0$ and
$|S_{z}|=1$ triplet components. It is seen that the current
$i_{3}$ due to the singlet component decreases with increasing
$\lambda _{h}$, and the currents $i_{0,1}$ due to the triplet
components increase with increasing $\lambda _{h}$. Interestingly,
the current $i_{1}$ caused by the triplet component with nonzero
projection of the total spin on the local $z$-axis is much larger
than the current $i_{0}$ caused by the $S_{z}=0$ triplet
component. Meanwhile the current $ i_{1}$ is absent in the tunnel
Hamiltonian method at all (compare Figs.1 and 2).

In Fig.3 we show the dependence of the total normalized critical
current $ i_{c}=i_{3}+i_{0}+i_{1}$ on $\lambda _{h}$ for $\theta
=0 $ on the normalized exchange field $\lambda_{h}$ (dotted line).
We compare
this dependence with the dependence $\bar{\imath}_{c}=\bar{\imath}_{3}+\bar{%
\imath}_{0}$ (solid line), i.e. with the critical current given by
the tunnel Hamiltonian
method, where $\bar{\imath}_{3,0}$ are determined by Eq.(\ref{JosCurrents103}%
) with $f_{3,0}^{2}(0)$ replaced by $\bar{f}_{3,0}^{2}$. One can
see a significant difference between these dependencies.

\bigskip

\section{Conclusions}

\bigskip We have studied the influence of boundary effects on properties of
magnetic superconductors with a spiral magnetic structure. We used
the well developed method of quasiclassical Green's functions.
These functions obey the Eilenberger (or Usadel) equations
generalized to the case of an exchange field $\mathbf{h}$ acting
on spins of free electrons and varying in space. For simplicity,
we considered the case of temperatures close to the critical one,
$T_{c}.$ Then, one can linearize equations for the condensate
matrix Green's functions $\check{f}.$ Due to a spatial dependence
of the exchange field $\mathbf{h}$, coefficients in the
Eilenberger (Usadel) equations depend on the coordinate $x$. We
excluded this dependence via a transformation which is equivalent
to introducing a rotating coordinate
system. In this local coordinate system the field $\mathbf{h}$ has only the $%
z$-component and does not depend on $x$. Solving these equations
with corresponding boundary conditions, we have shown that near
the boundary of \textit{MS} with vacuum or an insulator, the
condensate functions $\check{f}$ differ essentially from their
bulk values.

In the rotating coordinate system, there are two components of the matrix $%
\check{f},$ $\bar {f}_{3}$ and $\bar {f}_{0},$ in the bulk. These
correspond to the singlet component and the triplet component with
zero projection of the total spin on the $z$-axis. Due to boundary
conditions, the corrections $\delta f_{0,3}$ to the bulk
functions, $\bar{f}_{0,3}$, arise near the boundary, which are not
small in comparison with $\bar{f}_{0,3}.$ Besides, the triplet
component $F_{1}$ with nonzero projection of the total spin of
Cooper pairs appears in the vicinity of \ the surface on the scale
of the coherence length. The corrections $\delta f_{0,3}$ \ and
function $F_{1}$ oscillate with the period $2\pi /Q$ in space and
decay inside the bulk over
a length of the order of $\xi _{T}=v/2\pi T$ (ballistic case) or $\xi _{T}=%
\sqrt{D/2\pi T}$ (diffusive case). The amplitude of the singlet component $%
f_{3}$ decreases at the surface resulting in a suppression of the
order parameter $\Delta $ near the surface.

As an example of importance of the surface effects in
\textit{MS}s, we considered the dc Josephson effect in a
\textit{MS/I/MS} \ tunnel junction. The critical Josephson current
$I_{c}$ can be expressed in terms of components $f_{0,3}(0)$ and
$F_{1}(0)$ at the \textit{MS/I} interface. The results are
compared with the ones which are obtained on the basis of the
tunnel Hamiltonian method and expressed in terms of the bulk
condensate functions $\bar{f}_{0,3}$. This method was used in
Ref.\cite{Kulic}. Although the formulae for $I_{c}$ in
Ref.\cite{Kulic} and in this paper are similar, there is an
essential difference between them. In the tunnel Hamiltonian
method, the coefficient in front of $\cos \theta $ is the squared
amplitude of the triplet $S_{z}=0$ component, $\bar{f} _{0}^{2}$.
In fact, this coefficient is equal to $f_{0}^{2}(0)+F_{1}^{2}(0)$
(see Eq.(\ref{JosCurrent})), where $f_{0}(0)$ is the amplitude of
the triplet component with zero projection of the spin on the
local $z$-axis and $F_{1}(0)$ is the amplitude of the $|S_{z}|=1$
triplet component at the interface. It turns out that, at least
near $T_{c}$, the amplitude $F_{1}(0)$ is much larger than
$f_{0}(0).$ The tunnel Hamiltonian method can be applied to
\textit{MSs }only if the wave vector of the spiral, $Q$, is small
enough: $vQ<<h$ (ballistic case) or $DQ^{2}<<h$ (diffusive case).
However in this case the exchange field $h$ should be small:
$h<\Delta $ ($T<<\Delta $) or $h<<(T_{c}-T)/T$ ($\Delta <<T$).
Otherwise superconductivity will be destroyed. In this limit of
small $Q$, the junction \textit{MS/I/MS} is equivalent to the
\textit{FS/I/FS }junction.
The Josephson current in \textit{FS/I/FS} junctions was calculated in Refs.\cite%
{BVE01,Fominov01,Sudbo,Barash}.

The surface effects may also change other characteristics of
\textit{MS}s such as the density-of-states (DOS) etc. Our
consideration is restricted with temperatures $T$ near $T_{c}$,
where the DOS is close to that in the normal state and the
variation of the DOS due to the surface effects is small. The
calculation of the Green's functions in a finite system at low $T$
is a more complicated task because the corresponding equations,
strictly speaking, can not be linearized. This problem is beyond
the scope of this paper.

\section{\protect\bigskip Acknowledgements}

\bigskip
I would like to thank SFB 491 for financial support.

\section{Appendix}

Here we obtain a formula for the Josephson current $I_{J}$ in a \textit{%
MS/I/MS} tunnel junction. We consider magnetic superconductors
\textit{MS} with a spiral magnetization described by the angle
$\alpha (x)=Qx+\theta $
(right superconductor) and $\alpha (x)=Qx$ (left superconductor) so that $%
\theta $ is the angle between the magnetization vectors at the \textit{MS/I}
interface. In order to obtain the expression for $I_{J}$, we employ the
boundary conditions \cite{Zaitsev,Kupriyanov,BVErmp}

\begin{equation}
\check{f}_{l}\partial _{x}\check{f}_{l}\;=(2\sigma R_{B})^{-1}[\check{f}_{l},%
\check{f}_{r}]  \label{A1}
\end{equation}
where $\check{f}_{l,r}$ are the condensate functions in the left
(right) superconductor, $\sigma$ is the conductivity of the
superconductors in the normal state, and $R_{B}$ is the junction
resistance per unit area. The superconductors are assumed to be
identical. The current is equal to \cite{BVErmp}

\begin{equation}
I=(\mathcal{S}\sigma /8)i(2\pi T)\sum_{\omega }Tr\{\hat{\tau}_{3}\otimes
\hat{\sigma}_{0}\otimes \check{f}_{l}\otimes \partial _{x}\check{f}_{l}\}=%
\frac{\mathcal{S}}{16R_{B}}i(2\pi T)\sum_{\omega }Tr\{\hat{\tau}_{3}\otimes
\hat{\sigma}_{0}\otimes \lbrack \check{f}_{l},\check{f}_{r}]\}  \label{A2}
\end{equation}
where $\omega =\pi T(2n+1)$ is the Matsubara frequency and all the functions
are taken at the interface ($x=0$).

We assume that the phase of the left superconductor is $\varphi $
and the phase of the right superconductor is zero. Then, we can
express the functions $\check{f}_{l,r}$ in terms of the functions
$\check{f}$ found above with the help of transformations

\begin{equation}
\check{f}_{l}\Longrightarrow \check{U}_{\varphi }\mathbf{\otimes }\check{U}_{l}\mathbf{%
\otimes }\check{f}_{l}\mathbf{\otimes }\check{U}_{\varphi }^{+}\mathbf{%
\otimes }\check{U}_{l}^{+},\text{ }\check{f}_{r}\Longrightarrow \check{U}_{r}\mathbf{%
\otimes }\check{f}_{r}\mathbf{\otimes }\check{U}_{r}^{+}
\label{A3}
\end{equation}
Here $\check{U}_{\varphi }=\cos (\varphi /2)+i\hat{\tau}_{3}\otimes \hat{%
\sigma}_{0}\sin (\varphi /2)$ is the \ transformation matrix which relates a
state with phase equal to zero and a state with a finite phase $\varphi $ %
\cite{BVErmp}; $\check{U}_{l,r}=\cos (\alpha _{l,r}/2)+i\hat{\tau}%
_{3}\otimes \hat{\sigma}_{1}\sin (\alpha _{l,r}/2)$ with $\alpha
_{l}=Qx$ and $\alpha _{r}=Qx+\theta $. Then, we substitute
expressions (\ref{A3}) together with (\ref{fDif}) into
Eq.(\ref{A2}). Calculating the commutator in Eq.(\ref{A2}), we
come to Eq.(\ref{JosCurrent}).

It is worth noting that the tunnel Hamiltonian leads to the same
formula as Eq.(\ref{A2}) if the functions $\check{f}_{r,l}$ are
replaced by the bulk solutions, $\bar{f}_{0,3}$.


\begin{thebibliography}{99}
\bibitem{footnote} The parameter $\lambda _{\omega }(n)=(2\pi
T/DQ^{2})(2n+1)$ depends on $n$ with a characteristic $n_{ch}$
determined by a value at which corresponding sums over the
Matsubara frequencies converge. One can see from Eq.(\ref{fBar})
that $n_{ch}$ is of the order \ $(2n_{ch}+1)\gtrsim \lambda
_{h}(h/\pi T)$.
\end{thebibliography}
\end{document}